\documentclass[nofootinbib,11pt]{revtex4-2}
 
\usepackage{graphicx}
\usepackage[margin=1in]{geometry}
\usepackage{amsmath,amssymb,amsfonts,amsthm,stmaryrd,mathtools,bm}
\usepackage{physics}
\usepackage{mathrsfs}
\usepackage{color}
\usepackage{multirow,bigdelim}
\usepackage{caption}
\usepackage{subcaption}
\usepackage{hyperref}
\usepackage{cleveref}
\crefname{equation}{}{}
\usepackage{dsfont}
\usepackage{booktabs}
\allowdisplaybreaks[0]

\usepackage{pstricks}
\usepackage{tikz}
\usepackage{ulem}

\newcommand{\C}{\hat{\mathcal{C}}}

\newcommand{\J}[2]{J_{#1}\left(\frac{\sqrt{#2}}{\hbar}v\right)}

\def\be#1\ee{\begin{align}#1\end{align}}

\def\ba{\begin{eqnarray}}
\def\ea{\end{eqnarray}}

\newcommand{\oarX}[1]{\href{http://arxiv.org/abs/#1}{{\ttfamily #1}}}
\newcommand{\arX}[1]{\href{http://arxiv.org/abs/#1}{{\ttfamily arXiv:#1}}}

\begin{document}
\thispagestyle{empty}
\title{Unitarity and quantum resolution of gravitational singularities}
\author{Steffen Gielen}
\email{s.c.gielen@sheffield.ac.uk}
\affiliation{School of Mathematics and Statistics, University of Sheffield, Hicks Building, Hounsfield Road, Sheffield S3 7RH, United Kingdom}
\author{Luc\'ia Men\'endez-Pidal}
\email{lucia.menendez-pidal@nottingham.ac.uk}
\affiliation{School of Mathematical Sciences,   University of Nottingham, University Park, Nottingham NG7 2RD, United Kingdom}
\date{\today}

\begin{abstract}
We explore the consequences of requiring that quantum theories of gravity be unitary, mostly focusing on simple cosmological models to illustrate the main points. We show that unitarity for a clock that encounters a classical singularity at finite time implies quantum singularity resolution, but for a clock that encounters future infinity at finite time leads to a quantum recollapse. We then find that our starting point -- assuming the general covariance of general relativity -- is actually incompatible with general quantum unitarity: singularity resolution in quantum gravity can always be engineered by choosing the right clock, or avoided by using a different one. 
\end{abstract}

\maketitle

\vspace{5mm}
{\em Essay written for the Gravity Research Foundation 2022 Awards for Essays on Gravitation. Fifth Award}

\newpage
\setcounter{page}{0}

One of the most fascinating and troubling features of classical general relativity is its tendency to develop singularities from generic initial conditions, where the theory breaks down and must presumably be replaced by some quantum theory of gravity. Far from being just a mathematical curiosity, such singularities describe both the beginning of our Universe at the Big Bang and the endpoint of gravitational collapse in black holes. Understanding whether or how they might be resolved through quantum effects is hence one of the most important  questions of gravitational physics research.

We do not have a complete theory of quantum gravity, with fully understood foundations and unambiguous predictions, but the general consensus is that the principles of quantum theory should continue to apply when gravity is involved, lest we would not have a starting point for even formulating candidate theories. One of the most important properties of quantum theory is the {\em unitarity} of time evolution, which preserves the norm of the state:
\be
\langle \psi(t_1)|\psi(t_1) \rangle = \langle \psi(t_0)|\hat{U}^\dagger \hat{U}|\psi(t_0)\rangle = \langle \psi(t_0)|\psi(t_0)\rangle\,,
\label{unitarity}
\ee
where $\hat{U}$ is the unitary operator for time evolution from time $t_0$ to $t_1$. An immediate question in a generally covariant theory like general relativity is what time coordinate the ``$t$'' label in (\ref{unitarity}) should refer to; a natural starting point might be to demand that, unless we want to break general covariance, unitarity should hold for {\em all} possible choices of time coordinate.

Upon reflection, it becomes clear that this requirement is not physically meaningful; after all, values of a coordinate time are not observable. It does not make sense to think of a quantum state as a function of a coordinate label ``$t$''. This apparent absence of time evolution in quantum gravity has long puzzled researchers as the {\em problem of time} \cite{problemoftime}. While there are different approaches towards resolving it, the way out is the same as in classical relativity: meaningful physical statements must be made in coordinate-independent, {\em relational} terms. What {\em is} observable is the  answer to the question: ``what is the value taken by quantity $A$ when quantity $B$ satisfies $B=b_0$?,'' where $A$ and $B$ might refer to matter or gravitational degrees of freedom \cite{relationalobservables}. A quantity $B$ with useful dynamical properties might serve as a ``clock'' characterising the evolution of other degrees of freedom. Ideally, such $B$ evolves monotonically so that each ``$B=b_0$'' characterises exactly one instant of time, and its interaction with other degrees of freedom should be as weak as possible.

Our desideratum of unitarity in quantum gravity then becomes the demand that {\em time evolution should be unitary with respect to physically reasonable choices of clock degree of freedom.} When restricting to simple situations such as spherical symmetry or homogeneous cosmology, it is not too difficult to find models which contain one or more ideally suited clocks which should clearly be classed as ``physically reasonable,'' so that the notion of unitarity can be addressed explicitly. In a general framework without any symmetry, this would clearly be a more difficult problem but also there suitable matter clocks can be defined \cite{dustmod}.

So far, nothing we have said might seem particularly controversial. There is however an immediate clash between the notion of unitarity and the behaviour of classical general relativity: classical solutions tend to terminate in finite proper time along a geodesic, where they end in a singularity. Any model in which there is a physical (matter) clock measuring proper time will have to reconcile this classical behaviour with the requirement of unitarity, which says that the quantum state should be well-defined at all times: in particular, a state sharply peaked around some classical solution must be defined beyond the range where this classical solution is well-defined. In other words, our notion of {\em unitarity of quantum gravity is not compatible with the notion of singularities appearing after a finite time.} Singularities must then be resolved through quantum effects, which lead to departure from classical physics.

To the best of our knowledge, this argument was first presented as a general conjecture about quantum resolution of cosmological (spatially homogeneous) singularities by Gotay and Demaret in \cite{GotayDemaret}. They classified the possible clock variables near singularities as ``slow'' or ``fast'': a slow clock is one that only measures a finite amount of time before hitting the singularity, whereas a fast one measures an infinite amount. Any clock related to proper time is usually slow for matter satisfying the usual energy conditions, given that the classical singularity theorems tell us that singularities appear at finite proper time. An important subtlety is that the clock variable must be valued over the entire real line to avoid, e.g., redefinitions $T\rightarrow f(T)$ mapping the real line to an interval, which would make any clock variable $T$ appear ``slow''. As a well-known example consider ``Misner time'' $\log(v)$, where $v=\sqrt{\det h}$ is the volume element of space if $h_{ij}$ is the spatial metric. The cosmological singularity occurs at $\log(v)=-\infty$, and Misner time would be classified as a fast clock. In an expanding Universe, we might expect quantum gravity to be unitary with respect to Misner time and, as Misner already discusses in \cite{Misner}, this notion of unitarity is compatible with the persistence of the classical singularity in the quantum theory, given that this appears infinitely ``far'' in the past.  For slow clocks, however, unitarity is incompatible with the existence of a singularity. Hence, {\em whether classical singularities must be resolved in quantum gravity depends on the choice of clock}.

We have studied the implications of quantum-mechanical unitarity for the resolution of cosmological singularities in \cite{OurPapers}, confirming Gotay and Demaret's conjecture in a new example involving three possible choices of clock. Only one of the three clocks is slow with respect to the cosmological singularity, and only this choice of clock leads to a quantum theory that resolves the singularity by a ``bounce''. Such a bounce might be regarded as quite a desirable feature for a quantisation of gravity. However, even future infinity in an expanding Universe is only a finite time away for some choices of time. The example in our model is that of a free massless scalar field, which settles down to a constant value as the Universe expands. There is nothing unusual or unphysical about this clock, which is indeed a popular choice of matter clock in quantum cosmology. But if quantum gravity is unitary with respect to such a matter clock, {\em future infinity must be ``resolved'' by quantum effects just as we usually hope cosmological singularities to be}. The quantum state must continue beyond the time at which the classical solution terminates by reaching infinite volume; quantum effects trigger a quantum recollapse, or transition from future infinity to the past infinity of a new classical solution \cite{OurPapers}. The logic behind Gotay and Demaret's conjecture applies to future (or past) infinity just as it applies to singularities.

The cosmological model in \cite{OurPapers} is simple enough to be analytically solvable, but complicated enough to give insights into ambiguous physical consequences of demanding unitarity. It can be defined as a quantum spatially flat, homogeneous and isotropic Universe with metric
\be
{\rm d}s^2 = -N(\tau)^2{\rm d}\tau^2 + a(\tau)^2 h_{ij}\,{\rm d}x^i{\rm d}x^j
\ee
where $N$ is the lapse, $a$ the scale factor, and $h_{ij}$ is flat. This Universe is filled with a free, massless scalar field and subject to the dynamics of unimodular gravity \cite{unimodular} where the cosmological constant of general relativity appears as an integration constant. While classically equivalent to usual general relativity, this allows for superposition of different values of $\Lambda$ in the quantum theory, which can be used to define {\em one} notion of unitarity \cite{GrybThebault}.

The Hamiltonian for the dynamics of gravity and matter is given by
\be
H=N\left[-\frac{1}{12V_0a}\pi_a^2+\frac{1}{2V_0 a^3}\pi_\phi^2 + V_0 a^3\Lambda\right]
\label{ham-cc}
\ee
where we have chosen units in which $8\pi G =1$, $V_0=\int {\rm d}^3 x\sqrt{h}$ is the (unspecified) coordinate volume of space, and $\pi_a$ and $\pi_\phi$ are canonical momenta to the scale factor $a$ and scalar field $\phi$, respectively. $\Lambda$ is canonically conjugate to a dynamical variable $T$ (i.e., $\{T,\Lambda\}=1$) which plays the role of a Lagrange multiplier for the constraint ${\rm d}\Lambda/{\rm d}\tau=0$ \cite{OurPapers}. After suitable variable redefinitions (\ref{ham-cc}) can be brought into the simpler form
\be
H=\tilde{N}\left[-\pi_v^2 +\frac{\pi_\varphi^2}{v^2}+\lambda\right]
\label{hamnew}
\ee
where $\tilde{N}:=Na^3$ is a rescaled lapse and $v:=\sqrt{\frac{4V_0}{3}}a^3$ is proportional to the volume of space; $\pi_v$ is the momentum conjugate to $v$. Finally, $\lambda:=V_0\Lambda$ is a rescaled cosmological ``integration constant'' (now conjugate to $t:=V_0^{-1}T$) and $\pi_\varphi = C \pi_\phi$ for a suitable constant $C$ (so that $\varphi=C^{-1}\phi$ to obtain a canonical pair).

The dynamical equations arising from the classical Hamiltonian (\ref{hamnew}) can be solved analytically; they reproduce, of course, known cosmological solutions for a Universe with a free massless scalar field and a cosmological constant. In line with the philosophy we have outlined for quantum mechanical unitarity and to bring these solutions into a form that corresponds to observable (gauge-invariant) statements, the solutions must now be expressed in terms of an ``internal clock'' by choosing one of the dynamical variables as reference for the others. There are various choices one can make, but three obvious ones are the scalar field $\varphi$, the unimodular ``dark energy time'' $t$, and the Misner volume time $\log(v)$.

\begin{figure}
\centering
\begin{subfigure}{0.49\textwidth}
\includegraphics[width=\textwidth]{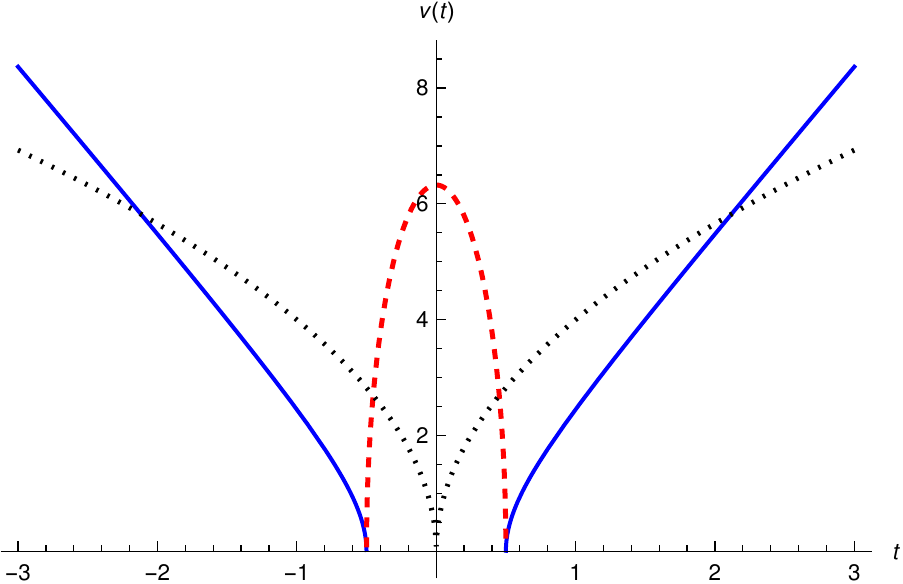}
\caption{$v(t)$ with singularity $v=0$ at finite $t$.}
\end{subfigure}
\begin{subfigure}{0.49\textwidth}
\includegraphics[width=\textwidth]{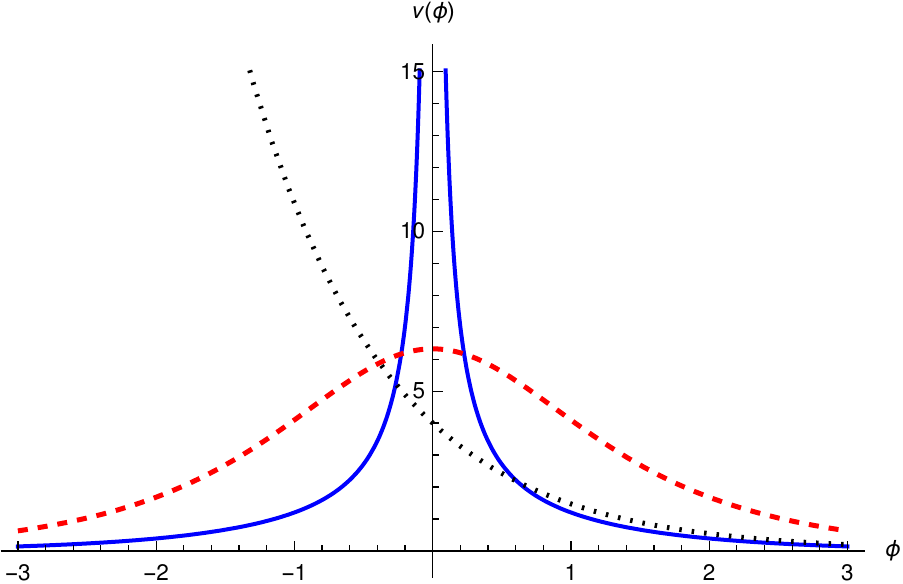}
\caption{$v(\varphi)$ with $v=\infty$ at finite $\varphi$ if $\lambda>0$.}
\end{subfigure}
\caption{Classical solutions $v(t)$ and $v(\varphi)$ for $\lambda>0$ (solid), $\lambda=0$ (dotted), $\lambda<0$ (dashed).}
\label{class-sol}
\end{figure}

We are interested in points where the volume $v$ reaches zero, corresponding to a big bang or big crunch singularity, or (past or future) infinity. In \cref{class-sol} we show the classical solutions for $v$ as a function of the clocks $t$ and $\varphi$. $t$ is a slow clock at the singularity and $\varphi$ is slow at infinity (for $\lambda>0$), which is where we expect a quantum departure from classical solutions.

We can quantise this model in the usual way by replacing canonical momenta by derivatives with respect to coordinates. There are ordering ambiguities which we fix by demanding covariance with respect to redefinitions of the $(v,\varphi)$ coordinates as in \cite{HawkingPage}; this leads to the Wheeler--DeWitt equation
\begin{equation}
\C\Psi(v,\varphi,t):=\left(\hbar^2\pdv[2]{}{v}+\frac{\hbar^2}{v}\pdv{}{v}-\frac{\hbar^2}{v^2}\pdv[2]{}{\varphi}-{\rm i}\hbar\pdv{}{t} \right)\Psi(v,\varphi,t)=0\, .
\label{wdw}
\end{equation}
Given that the classical Hamiltonian $H$ must vanish, $\Psi$ must be annihilated by its quantum analogue $\C$. The naive time evolution operator in \cref{unitarity}, given by $e^{-{\rm i} \C \tau}$ for some time evolution parameter $\tau$, would be equal to the identity on any physical state, which is another way of describing the problem of time. The way out, as we have stressed before, is to choose one of $\log(v)$, $\varphi$ or $t$ as clock and view (\ref{wdw}) as describing evolution of $\Psi$ {\em in this particular} time: this transforms the problem of time into a {\em multiple choice problem}.

The general solution to (\ref{wdw}) can be given in terms of Bessel functions,
\begin{align}
\Psi(v,\varphi,t)=&\int_{-\infty}^\infty \frac{\dd \lambda}{2\pi \hbar} \int_{-\infty}^\infty\frac{\dd k}{2\pi}e^{{\rm i}k\varphi}e^{{\rm i}\lambda \frac{t}{\hbar}}\left[\alpha(k,\lambda)\J{{\rm i}\abs{k}}{\lambda} + \beta(k,\lambda)\J{-{\rm i}\abs{k}}{\lambda}\right]\nonumber\\
+&\int_{-\infty}^{\infty} \frac{\dd \lambda}{2\pi \hbar} \int_{-\infty}^\infty\frac{\dd \kappa}{2\pi}e^{\kappa\varphi}e^{{\rm i}\lambda \frac{t}{\hbar}}\left[\gamma(\kappa,\lambda)\J{\abs{\kappa}}{\lambda} + \epsilon(\kappa,\lambda)\J{-\abs{\kappa}}{\lambda}\right]\, .
\label{general-sol}
\end{align}

The crucial point is now to demand that the quantum theory be unitary, i.e., that a suitably chosen inner product is conserved under ``time'' evolution for our three possible physical clocks. Writing the Wheeler--DeWitt equation for our three clock choices as 
\begin{equation}
{\rm i}\hbar\pdv{}{t}\Psi=\C_1\Psi, \quad-\hbar^2\pdv[2]{}{\varphi} \Psi=\C_2\Psi, \quad-\hbar^2\pdv[2]{}{\log(v)}\Psi=\C_3\Psi
\label{schr-kg-eq}
\end{equation}
where
\be
\C_1=\hbar^2\pdv[2]{}{v}+\frac{\hbar^2}{v}\pdv{v}-\frac{\hbar^2}{v^2}\pdv[2]{}{\varphi}\,, \quad \C_2=-\hbar^2\left(v\pdv{}{v} \right)^2+{\rm i}\hbar v^2\pdv{}{t}\,, \quad
\C_3=-\hbar^2\pdv[2]{}{\varphi}-{\rm i}\hbar v^2\pdv{}{t}\,,
\ee
the natural requirement is to have unitarity in a Schr\"odinger inner product for the $t$ clock, and in a Klein--Gordon inner product if $\varphi$ or $\log(v)$ is the clock. Each of these conditions translates into a requirement for $\C_1$, $\C_2$ or $\C_3$ to be self-adjoint in a corresponding $L^2$ inner product. One finds that while $\C_3$ is time-dependent (as it depends on $v$) but automatically self-adjoint, demanding self-adjointness of $\C_1$ or $\C_2$ leads to a boundary condition at $v=0$ for $\C_1$ and at $v=\infty$ for $\C_2$; at these points these operators are equivalent to strongly attractive potentials in standard quantum mechanics. The boundary conditions then lead to a reflection of the quantum state at points where the classical solutions would terminate (see again \cref{class-sol}), explicitly realising the general argument we outlined above. The quantum state, even if it was initially sharply peaked around a classical trajectory, cannot follow the classical solution into its singularity or into infinity; it must be reflected and mapped into a different solution with opposite arrow of time.

The consequences of this behaviour can be understood most directly when computing quantum expectation values in these theories. As analogues of the classical solutions $v(t)$ and $v(\varphi)$, we can compute $\bra{\Psi_{sc}(t)}v\ket{\Psi_{sc}(t)}=\expval{v(t)}_{\Psi_{sc}}$ for the Schr\"odinger-like $t$ theory and $\bra{\Psi_{sc}(\varphi)}v\ket{\Psi_{sc}(\varphi)}=\expval{v(\varphi)}_{\Psi_{sc}}$ for the Klein--Gordon-like $\varphi$ theory, where $\Psi_{sc}$ is a semiclassical state obtained by choosing sharply peaked Gaussians as $\alpha(k,\lambda)$ and $\beta(k,\lambda)$ in \cref{general-sol}. For Gaussians peaked around $k=k_c$ and $\lambda=\lambda_c$, these expectation values can be compared to classical trajectories $v(t)$ and $v(\varphi)$ where $\lambda=\lambda_c$ and $\pi_\varphi=\hbar k_c$. If $\log(v)$ is the clock, $v$ itself is no longer an observable but we can compare $\bra{\Psi_{sc}(v)}t\ket{\Psi_{sc}(v)}=\expval{t(v)}_{\Psi_{sc}}$ to a classical trajectory $t(v)$. 

\begin{figure}
\centering
\begin{subfigure}{0.45\textwidth}
\centering
\includegraphics[width=\textwidth]{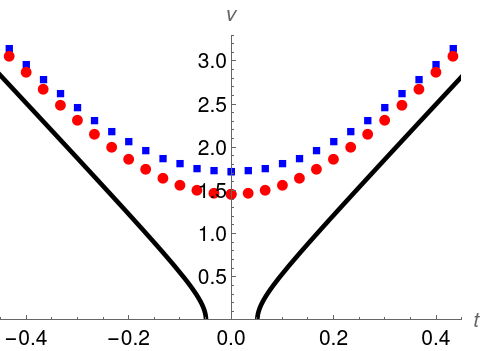}
\caption{$\expval{v(t)}_{\Psi_{sc}}$ compared to $v(t)$ (bold)}
\label{vt-fig}
\end{subfigure}
\begin{subfigure}{0.49\textwidth}
\centering
\includegraphics[width=\textwidth]{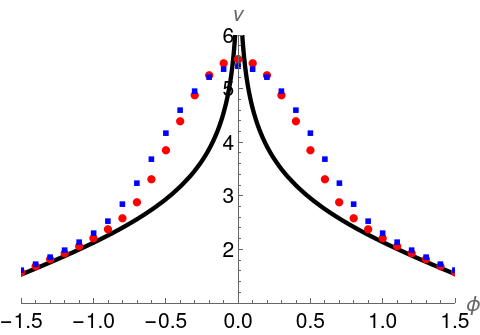}
\caption{$\log\expval{v(\varphi)}_{\Psi_{sc}}$ compared to  $\log v(\varphi)$ (bold)}
\label{vphi-fig}
\end{subfigure}
\begin{subfigure}{0.5\textwidth}
\includegraphics[width=\textwidth]{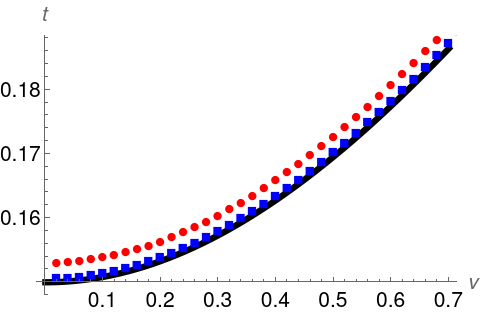}
\caption{$\expval{t(v)}_{\Psi_{sc}}$ compared to $t(v)$ (bold)}
\label{tv-fig}
\end{subfigure}
\caption{Expectation values in three different unitary quantum theories.}
\label{dynamics}
\end{figure}

\Cref{dynamics} shows how classical singularities and spatial infinity are both ``resolved'' by unitary quantum dynamics. In \cref{vt-fig}, the singularity is replaced by a big bounce (the expectation value of $v$ remains bounded away from zero), and in \cref{vphi-fig} we see how the Universe reaches a maximum volume and then recollapses due to quantum effects. \Cref{tv-fig} shows that in the Misner clock theory expectation values do not depart from the classical trajectories. The different colours in these figures correspond to different choices of Gaussian.

Mathematically, the behaviour seen in our model is easy to understand: quantum evolution must extend beyond the ``time'' when a classical solution would terminate, and this extension removes classical singularities or can replace infinity by a recollapse. More specifically we saw that this behaviour results from boundary conditions needed to have self-adjointness of an operator appearing in the corresponding time evolution problem.

However, the implications are now rather startling. It looks like the old dream of resolving classical singularities in quantum gravity is easy to achieve: all we need to do is pick a quantum clock that is slow where we want a singularity to be resolved, and demand that quantum theory remains unitary. We showed this explicitly in a simple homogeneous cosmological model, but there is no reason why similar constructions would not be possible for black holes or other spacetimes of physical interest. Black hole singularities would then also be replaced, presumably by a ``bounce'' into a white hole, as is often expected in quantum gravity. No departure from general relativity as the theory of gravity would be needed, and no introduction of new quantisation principles or degrees of freedom beyond the spacetime metric, as for instance in string theory or loop quantum gravity.

At the same time, we are abandoning the main principles of general relativity in defining such quantum theories. The meaning of unitarity is tied to a specific clock choice and has different physical consequences for different clocks, in stark contrast with the general covariance of classical general relativity that we started off with. As another drastic example of what we seem to have concluded, notice that de Sitter space can be foliated by positively curved slices leading to a scale factor that is globally regular, $a(t)\sim \cosh(\sqrt{\Lambda/3}\,t)$, or by negatively curved slices with $a(t)\sim \sinh(\sqrt{\Lambda/3}\,t)$, with a coordinate ``singularity'' at $t=0$. The same general arguments would now suggest that, in an appropriate time coordinate such as proper time, the open slicing could require ``singularity resolution'' at $t=0$ where classical solutions terminate, and quantum evolution would necessarily depart from the classical solution; but the closed slicing is classically well-defined and does not require any quantum corrections. In this example {\em the same classical spacetime} has two different quantum analogues depending on the foliation, again a violent breaking of general covariance.

If we decide that general covariance is sacred in quantum gravity, it seems that it is the notion of unitarity that has to go, in the sense that it might apply only to some clocks but not others, even if these are equally good clocks classically. How are we to select these preferred notions of time and what singles them out? What interpretation of quantum mechanics are we supposed to attach to a time in which the ``sum of probabilities'' is not conserved? Our attempt at working from a conservative starting point -- only relying on basic principles of general covariance in relativity and unitarity of quantum mechanics -- has clearly failed, as it has led us to the conclusion that some basic principles need to be abandoned after all.

{\bf Acknowledgments.} --- The work of SG was funded by the Royal Society through a University Research Fellowship (UF160622) and a Research Grant for Research Fellows (RGF\textbackslash R1\textbackslash 180030).

\end{document}